\documentclass[12pt]{article}
\usepackage{epsfig}
\usepackage{amsmath}
\usepackage{hhline}
\usepackage{amssymb}
\usepackage{times}

\newlength{\dinwidth}
\newlength{\dinmargin}
\newcommand{\GeV}{\rm GeV}
\newcommand{\TeV}{\rm TeV}
\newcommand{\pb}{\rm pb}
\newcommand{\cm}{\rm cm}
\newcommand{\hdick}{\noalign{\hrule height1.4pt}}

\begin{document}

\pagestyle{empty}
\begin{titlepage}
\begin{flushleft}
  {\tt DESY 00-027} \hfill {\tt ISSN 0418-9833} \\
  {\tt February 2000}  
\end{flushleft}

\vspace*{3cm}

\begin{center}
  \LARGE
  {\bf Search for Compositeness, Leptoquarks \\ 
    and Large Extra Dimensions \\ 
    in {\boldmath $e q$} Contact Interactions at HERA}

  \vspace*{1cm}
    {\Large H1 Collaboration} 
\end{center}
\normalsize
\vspace*{2cm}

\begin{abstract}
  \noindent
  The reaction $e^+p \rightarrow e^+X$ is studied with the H1 detector 
  at {\sc Hera}. The data cover momentum transfers $Q^2$ between 
  $200~\GeV^2$ and $30,000~\GeV^2$ and correspond to an integrated 
  luminosity of $35.6~\pb^{-1}$.
  The differential cross section ${\rm d}\sigma / {\rm d} Q^2$ is compared
  to the Standard Model expectation for neutral current scattering
  and analysed to search for $(\bar{e} e)(\bar{q} q)$ contact interactions.
  No evidence for new phenomena is observed.
  The results are used to set limits on scales within models of 
  electron--quark compositeness, quark form factors and the exchange of
  virtual heavy leptoquarks.
  A search for gravitational effects mediated through the exchange of
  virtual gravitons which propagate into large extra dimensions is presented. 
\end{abstract}

\vspace*{2.0cm}
\begin{center}
  {\it Submitted to Physics~Letters~B}
\end{center}

\vfill

\newpage
\begin{flushleft}
 C.~Adloff$^{33}$,                
 V.~Andreev$^{24}$,               
 B.~Andrieu$^{27}$,               
 V.~Arkadov$^{35}$,               
 A.~Astvatsatourov$^{35}$,        
 I.~Ayyaz$^{28}$,                 
 A.~Babaev$^{23}$,                
 J.~B\"ahr$^{35}$,                
 P.~Baranov$^{24}$,               
 E.~Barrelet$^{28}$
% ...

  \\[5ex] {\it 
% ...
$ ^{23}$ Institute for Theoretical and Experimental Physics,
          Moscow, Russia \\
 $ ^{24}$ Lebedev Physical Institute, Moscow, Russia$^{e,h}$ \\
% ...
$ ^{27}$ LPNHE, \'{E}cole Polytechnique, IN2P3-CNRS, Palaiseau, France \\
 $ ^{28}$ LPNHE, Universit\'{e}s Paris VI and VII, IN2P3-CNRS,
          Paris, France \\
% ...
$ ^{33}$ Fachbereich Physik, Bergische Universit\"at Gesamthochschule
          Wuppertal, Wuppertal, Germany$^a$ \\
% ...
$ ^{35}$ DESY, Zeuthen, Germany$^a$ \\
% ...
$ ^a$ Supported by the Bundesministerium f\"ur Bildung, Wissenschaft,
        Forschung und Technologie, FRG,
        under contract numbers 7AC17P, 7AC47P, 7DO55P, 7HH17I, 7HH27P,
        7HD17P, 7HD27P, 7KI17I, 6MP17I and 7WT87P \\
% ...
$ ^e$ Supported by the Deutsche Forschungsgemeinschaft \\
% ...
$ ^h$ Supported by Russian Foundation for Basic Research 
       grant no. 96-02-00019 \\
% ...
 }
\end{flushleft}

\end{titlepage}

\pagestyle{plain}


\section{Introduction \label{introduction}}

The {\sc Hera} collider enables the study of deep inelastic neutral current 
scattering $e p \rightarrow e X$ at very high squared momentum transfers $Q^2$,
thus probing the structure of $e q$ interactions at very short distances.
At large scales new phenomena {\em not directly} detectable 
may become observable as deviations from the Standard Model predictions.
A convenient tool to assess the experimental sensitivity beyond the 
maximal available center of mass energy
and to parameterise indirect signatures 
of new physics is the concept of four-fermion contact interactions.
Possible sources of such contact terms are either a substructure of the 
fermions involved or the exchange of a new heavy particle.
In the first case a compositeness scale can be related to the size of the
composite object, while in the second case the scale parameter is related
to the mass and coupling of the exchanged boson.

The principle idea of this contact interaction analysis at {\sc Hera} is to fix
the Standard Model and its parameters, in particular the parton distributions,
using experimental data at low $Q^2$, 
where the theory is well established, 
and to extrapolate the prediction towards high momentum transfers, 
where deviations due to new physics are expected to be most prominent.
In the present paper the differential cross section 
${\rm d}\sigma/{\rm d}Q^2$ is analysed 
over a $Q^2$ range of $200 - 30,000~\GeV^2$
and possible deviations from the Standard Model prediction are searched for 
in the framework of $(\bar{e} e)(\bar{q} q)$ contact interactions.
The data are interpreted within conventional scenarios such as 
model independent compositeness scales of various chiral structures,
a classical quark form factor approach and the exchange of heavy leptoquarks.
Another investigation concerns the search for low scale quantum gravity 
effects, which may be observable at {\sc Hera} via the exchange of gravitons 
coupling to  Standard Model particles and propagating into extra spatial
dimensions.


\section{Data Analysis \label{analysis}}

\begin{figure}[p] 
  \begin{center}
    \epsfig{file=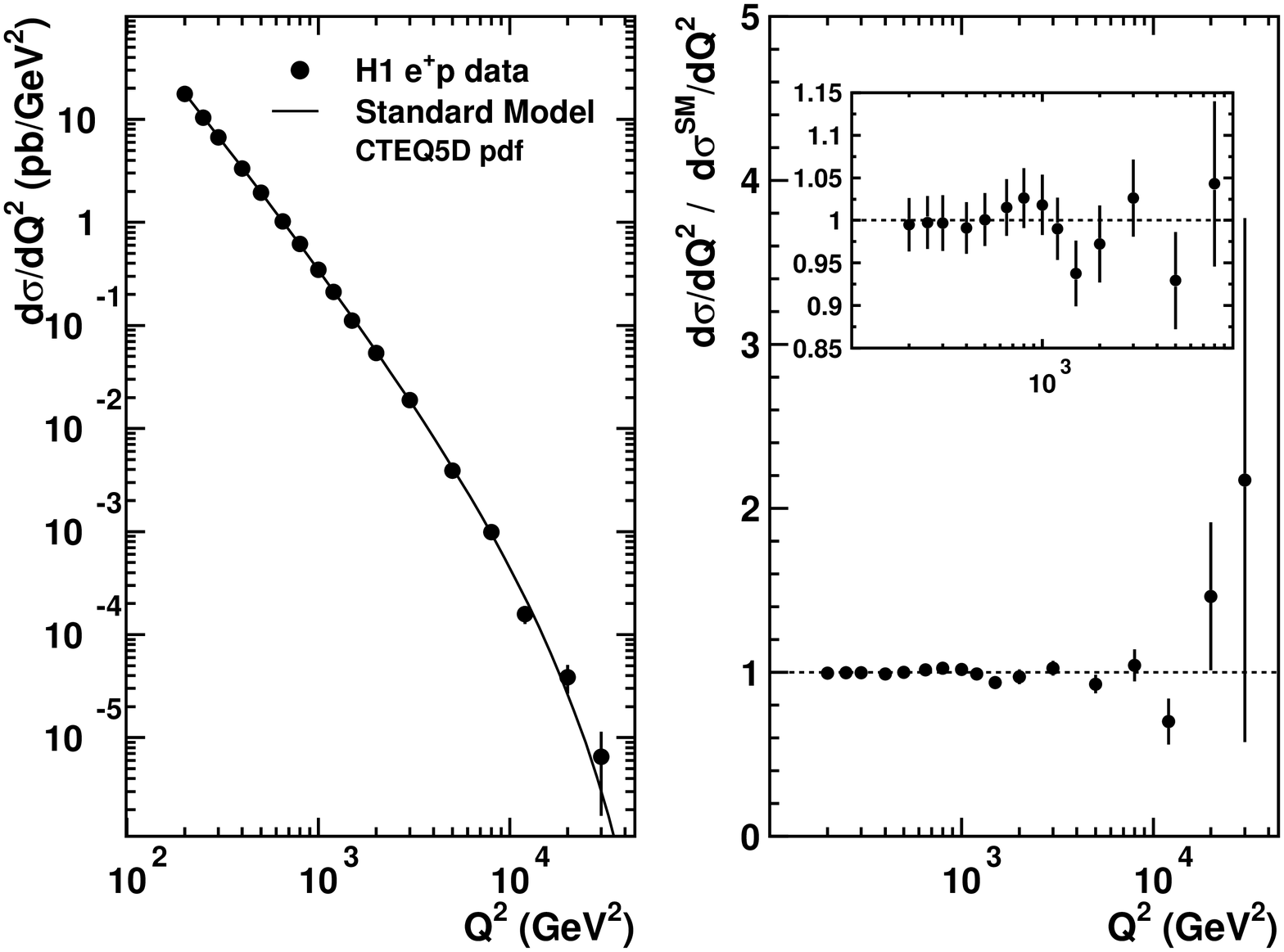,width=16cm}
  \end{center}
  \caption{Differential NC cross section 
    ${\rm d}\sigma (e^+p \rightarrow e^+X) / {\rm d}Q^2$.
    H1 data ($\bullet$) are compared with the Standard Model expectation
    (---) using CTEQ5D parton distributions.
    The errors represent statistics and uncorrelated experimental systematics. 
    The overall normalisation uncertainty is 1.5\%.}
  \label{cismxsec}
\end{figure}

The contact interaction analysis is based on the recent 
$e^+p$ neutral current cross section measurements by the H1 experiment
discussed in detail in ref.~\cite{h1xsec}. 
The data have been collected at a center of mass energy of 
$\sqrt{s} = 300~\GeV$ and
correspond to an integrated lumino\-sity of ${\cal L} = 35.6~\pb^{-1}$,
representing a tenfold increase over a previous analysis~\cite{h1ci}.
The cross section ${\rm d}\sigma/{\rm d}Q^2$
is determined from a purely inclusive measurement of the
final state positron with energy $E'_e$ and polar angle $\theta_e$
(defined with respect to the proton direction).
The squared momentum transfer is calculated via
$Q^2 = 4\,E_e\,E'_e\,\cos^2(\theta_e/2)$, 
where $E_e$ is the lepton beam energy.
The data are corrected for detector effects and QED radiation
and represent the cross section within the kinematic phase space of
momentum transfer $Q^2 \geq 200~\GeV^2$ and inelasticity 
$y = 1-E'_e/E_e\,\sin^2(\theta_e/2) < 0.9$.
The dominant experimental systematics 
are the uncertainties of the reconstructed positron energy scale, 
varying between $0.7\%$ and $3\%$ depending on the detector region,
and of the scattering angle, known to $1-3$~mrad.
An overall normalisation uncertainty of $1.5\%$ is due to the luminosity
determination.
The differential cross section is shown in figure~\ref{cismxsec}.

The double differential cross section is given in the
Standard Model by
\begin{eqnarray}
  \frac{\mathrm{d}^2\sigma(e^+ p \rightarrow e^+ X)}
       {\mathrm{d}x\,\mathrm{d}Q^2} & = & 
  \frac{2\,\pi\,\alpha^2}{x\,Q^4}\,
  \left\{ \, Y_+\,F_2(x,Q^2) 
    - Y_-\,x\,F_3(x,Q^2) - y^2\,F_L(x,Q^2) \, \right\}\ , 
  \label{xsection}
\end{eqnarray}
where $x = Q^2/y\,s$ is the Bjorken scaling variable and 
$Y_\pm = 1 \pm (1 - y)^2$.
The generalised proton structure functions 
$F_2(x,Q^2)$, $F_3(x,Q^2)$ and $F_L(x,Q^2)$
are related to the parton densities and the quark-$\gamma$ and 
quark-$Z$ couplings.
The cross section  calculations are done in the DIS scheme in 
next-to-leading-order using as standard the CTEQ5D parton 
parameterisation~\cite{cteq}.
Integrating eq.~(\ref{xsection}) over $x$ gives the $Q^2$ spectrum
which describes the data very well over six orders of magnitude,
see figure~\ref{cismxsec}.

In order to derive quantitative tests of the Standard Model and to search 
for new physics hypotheses, a $\chi^2$ analysis of the data is performed
taking the dominant error sources and uncertainties into account. 
The $\chi^2$ function is defined as
\begin{eqnarray}
  \chi^2 & = & \sum_{i} \, \left (
               \frac{\hat\sigma^{exp}_i f_n - \hat\sigma^{th}_i
                    \, (1 -  \sum_k \Delta_{i k}(\varepsilon_k)) }
                    {\Delta \hat\sigma^{exp}_i  f_n}
               \right )^2
               + \left(\frac{f_n-1}{\Delta f_n} \right )^2 
               + \sum_{k} \, \varepsilon^2_k \ .
  \label{chi2fcn}
\end{eqnarray}
Here $\hat\sigma_i$ denotes the experimental or theoretical cross section in
the $Q^2$ bin $i$ and $f_n$ is the overall normalisation parameter
with an uncertainty $\Delta f_n = 0.015$. 
The experimental error $\Delta \hat\sigma^{exp}_i$ includes
statistical and uncorrelated systematic errors added in quadrature.
The functions $\Delta_{i k}(\varepsilon_k)$ describe for the $i^{th}$ bin 
effects due to correlated systematic errors associated to different
sources $k$. 
They depend quadratically on the fit parameters $\varepsilon_k$, 
which may be interpreted as pulls, {\em i.e.} shifts caused by systematics
normalised to their error estimates.
There are three sources of correlated systematic errors taken into account:
the experimental uncertainties of the positron energy scale and the 
scattering angle and the uncertainty of the strong coupling entering in the
Standard Model prediction (see below).

Concerning cross section calculations the major uncertainty comes from
the parton distributions, which are generally provided without error estimates.
Different parametrisations in the DIS scheme,
MRST~99~\cite{mrst} and GRV~94~\cite{grv} in addition to CTEQ5D,
are used to estimate the uncertainties due to various models and assumptions.
They do not differ in the shape of the $Q^2$ spectrum significantly,
but rather in the absolute cross section prediction by up to 2.8\%,
{\em e.g.} comparing CTEQ5D with MRST~99.
Several other MRST sets are used for cross checks, like those with different
admixtures of quarks and gluons at high $x$, or different treatments
of {\em strange} and {\em charm} quarks.
All these MRST variants essentially change the overall normalisation of the cross
section prediction by less than $1\%$, being well below the measurement
errors, and introduce no relevant additional $Q^2$ dependence.
The largest uncertainty comes from the strong coupling constant.
Using parton distributions evaluated for couplings differing from the 
central value of $\alpha_s(M_Z) = 0.118$ by $\pm 0.005$ 
cause variations of the cross section by $\pm 1\%$ at low $Q^2$ and
up to $\mp 4\%$ at high $Q^2$.
These shifts are parameterised and taken into account as correlated 
systematic error 
in the $\chi^2$ fit of eq.~(\ref{chi2fcn}).
It should be noted that the applied parton density functions have not been
constrained by high $Q^2$ data from the {\sc Hera} experiments.
A comparison with a recent QCD analysis in the 
$\overline{\mbox{MS}}$ scheme~\cite{botje}, 
which attempts to provide parton distributions including errors, 
confirms that the above choice of various parton density functions is a 
reasonable representation of systematic uncertainties.

A fit of the cross section ${\rm d}\sigma/{\rm d}Q^2$ to the Standard Model 
expectation using CTEQ5D parton densities yields 
$\chi^2/{\rm dof} = 12.3/16$ with a normalisation parameter $f_n=1.004$.
Limits of a model parameter are derived by varying the parameter until the
$\chi^2$ value changes by a certain amount with respect to the Standard Model
fit, {\em e.g.} $\chi^2 - \chi^2_{SM} = 3.84$ for 95\% confidence level (CL).
Systematics due to different parton distributions are taken into account
by always quoting the most conservative result of the various fits, 
{\em i.e.} the smallest value in case of a lower limit.


\section{Contact Interaction Phenomenology \label{phenomenology}}  

New currents or heavy bosons may produce indirect effects
through the exchange of a virtual particle interfering with
the $\gamma$ and $Z$ fields of the Standard Model.
For particle masses and scales well above the available energy, 
$\Lambda \gg \sqrt{s}$,
such indirect signatures may be investigated by searching for 
a four-fermion pointlike $(\bar{e}\,e)(\bar{q}\,q)$ contact interaction. 
The most general chiral invariant 
Lagrangian for neutral current vector-like contact interactions
can be written in the form~\cite{elpr,haberl}
\begin{eqnarray}
  {\cal L}_V  &=&\sum_{q \, = \, u,\, d}\left\{\eta^q_{LL}\,
   (\bar{e}_L\gamma_\mu e_L)(\bar{q}_L\gamma^\mu q_L)
   +\eta^q_{LR}\,
   (\bar{e}_L\gamma_\mu e_L)(\bar{q}_R\gamma^\mu q_R)\right.
 \nonumber \\
   &&\ \ \ \ \left.+\;\eta^q_{RL}\,(\bar{e}_R\gamma_\mu e_R)
   (\bar{q}_L\gamma^\mu q_L) +\eta^q_{RR}\,
   (\bar{e}_R\gamma_\mu e_R)(\bar{q}_R\gamma^\mu q_R)\right\} \; ,
 \label{lcontact}
\end{eqnarray}
where the indices $L$ and $R$ denote the left-handed and right-handed
fermion helicities and the sum extends over {\em up}-type and 
{\em down-type} quarks and antiquarks $q$.
In deep inelastic scattering at high $Q^2$ the contributions from the
first generation $u$ and $d$ quarks completely dominate and contact terms 
arising from sea quarks $s$, $c$ and $b$ are strongly suppressed.
Thus, there are eight independent effective coupling coefficients,
four for each quark flavour 
\begin{eqnarray}
  \eta_{ab}^q & \equiv & \epsilon\frac{g^2}{\Lambda^{q \ 2}_{ab}} \ ,
  \label{etacoeff}
\end{eqnarray}
where $a$ and $b$ indicate the $L,\ R$ helicities,
$g$ is the overall coupling strength, $\Lambda^q_{ab}$ is a scale parameter
and $\epsilon$ is a prefactor, often set to $\epsilon = \pm 1$, 
which determines the interference sign with the Standard Model currents.
The ansatz eq.~(\ref{lcontact}) can be easily applied to any new 
phenomenon, {\em e.g.} $(e q)$ compositeness, leptoquarks or new gauge bosons, 
by an appropriate choice of the coefficients $\eta_{ab}$.
Scalar and tensor interactions of dimension~6 operators
involving helicity flip couplings  are strongly suppressed 
at {\sc Hera}~\cite{haberl} and therefore not considered.

It has been recently suggested that gravitational effects may
become strong at subatomic distances and thus measurable in collider
experiments~\cite{add}.
In such a scenario, which may be realised in string theory,
gravity is characterised by a scale $M_S \sim {\cal O}(\TeV)$ in $4+n$ 
dimensions.
The extra spatial dimensions $n$ are restricted to a volume associated with 
the size $R$ and the scales in $4+n$ and the ordinary $4$ dimensions are 
related by
\begin{eqnarray}
  M_P^2 & \sim & R^n \, M_S^{2+n} \ ,
 \label{mscale}
\end{eqnarray}
where $M_P \sim 10^{19}~\GeV$ is the Planck mass. 
An exciting consequence would be a modification of Newton's law at
distances $r < R$, where the gravi\-tational force would rise rapidly 
as $F \propto 1/r^{2+n}$ and become strong at the scale $M_S$.
Experimentally, gravity is essentially not tested in the 
sub-millimeter range~\cite{long} and scenarios with $n > 2$ extra dimensions 
at large distances $R \lesssim 100~\mu$m are conceivable.

In the phenomenology of~\cite{giudice} the Standard Model particles are 
confined to 4 dimensions while only the graviton propagates as massless spin 2 
particle into the $n$ extra dimensions.
When projected onto the normal 4 dimensional space the graviton appears as a 
spectrum of Kaluza-Klein excitations with masses $m^{(j)} = j/R$, 
including the zero-mass state.
The graviton fields $G^{(j)}_{\mu\nu}$ couple to the Standard Model particles 
via the energy-momentum tensor $T_{\mu\nu}$ 
\begin{eqnarray}
  {\cal L}_G & = & -\frac{\sqrt{8\,\pi}}{M_P} \,
                   G^{(j)}_{\mu\nu} \, T^{\mu\nu} \ .
 \label{lgraviton}
\end{eqnarray}
Summation over the whole tower of Kaluza-Klein states $j$ with masses
up to the scale $M_S$ 
compensates the huge $1/M_P$ suppression and
results in an effective contact interaction coupling 
\begin{eqnarray}
  \eta_G & = & \frac{\lambda}{M_S^4} \ ,
 \label{gscale}
\end{eqnarray}
where $\lambda$ is the coupling strength of order unity. 
Note that the scale dependence of gravitational effects is very different from
that of conventional contact interactions, eq.~(\ref{etacoeff}).
In deep inelastic scattering graviton exchange may contribute to the
electron-quark subprocess, but the new interaction also induces
electron-gluon scattering which is not present in the Standard Model.

It is worth recalling that contact interactions as an effective theory 
can only be formulated in lowest order.
They contribute~\cite{haberl} to the structure functions $F_2(x,Q^2)$ and 
$x F_3(x,Q^2)$, but are absent in $F_L(x,Q^2)$.
On the other hand a cross section calculation in next-to-leading-order QCD 
appears to be more reliable.
This conceptual limitation is less important in the
DIS renormalisation scheme, where the expression for the dominant
structure function $F_2$ is identical and $x F_3$ receives only minor 
corrections in second order.

Contact interaction phenomena are best observed as a modification of the
expected $Q^2$ dependence and all information is essentially contained
in the differential cross section $\mathrm d \sigma / \mathrm d Q^2$,
analysed in the present paper.
Calculations using the Standard Model prediction, eq.~(\ref{xsection}),
show that for the scenarios under study
only those models involving both $u$ and $d$ quarks with pure $LL$ or $RR$ 
couplings and negative interference are slightly more sensitive
to an ana\-lysis in two variables $Q^2$ and $x$.
With the present luminosity the gain 
in setting limits on the respective scales would be $\sim20\%$ for the 
negatively interfering $LL$ and $RR$ composite models 
and $\sim 10\%$ for the leptoquark $S^L_1$.
For all other scenarios the differences are marginal.


\section{Compositeness Scales}

In the Standard Model the fundamental particles -- leptons, quarks and gauge
bosons -- are assumed to be pointlike.
A possible fermion compositeness or substructure can be expressed through
the $\eta$ coefficients of eq.~(\ref{etacoeff}) which depend only on the
ratio of the coupling constant over the scale.
In the present analysis the interference sign is set to $\epsilon = \pm 1$
for the chiral structures under study,
the coupling strength $g$ is by convention chosen as $g^2 = 4\,\pi$ and 
the compositeness scale $\Lambda$ is assumed to be the same for all 
{\em up}-type and {\em down}-type quarks.
Among the many possible combinations the following models are investigated:
(i) the pure chiral couplings $LL$, $LR$, $RL$ and $RR$,
(ii) the vectorial couplings $VV$, $AA$ and $VA$,
(iii) the mixtures $LL + RR$ and $LR + RL$.

\begin{figure}[p] 
  \begin{center}  
    \epsfig{file=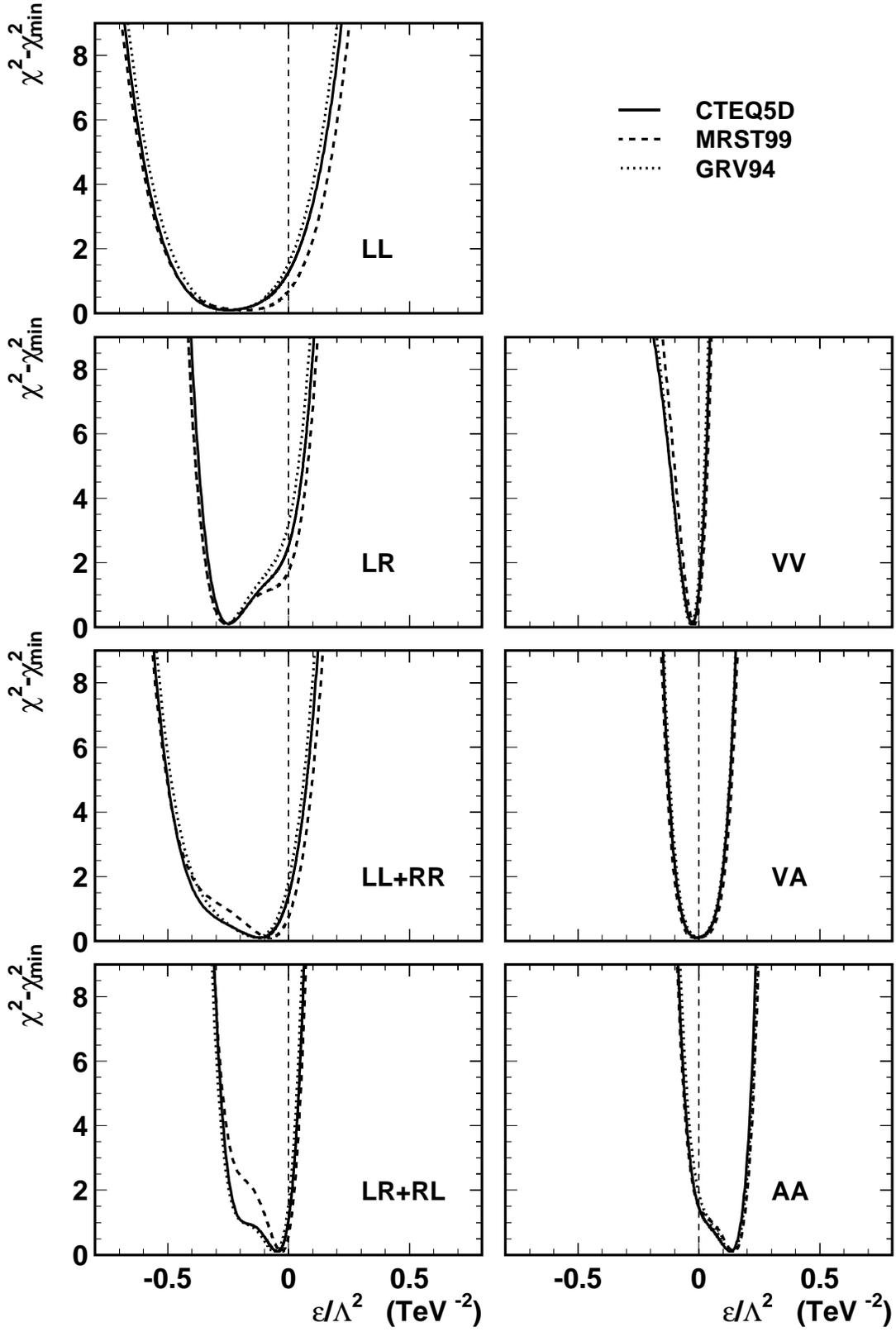,%
      bbllx=0pt,bblly=0pt,bburx=480pt,bbury=660pt,clip=,%
      width=16cm}
  \end{center}  \vspace{-.5cm}
  \caption{Distributions of $\chi^2- \chi^2_{min}$ versus $\epsilon/\Lambda^2$ 
    from fits to various compositeness models using CTEQ5D, MRST~99 and GRV~94 
    parton distributions and including full error propagation.}
  \label{cichi2eta}
\end{figure}

It is appropriate to analyse the differential cross section in terms of 
the coupling coefficients $\eta = \epsilon\,4\,\pi/\Lambda^2$.
Figure~\ref{cichi2eta} shows the values of $\chi^2$  
as a function of $\epsilon/\Lambda^2$ from fits to the models under study.
In general one observes that the distributions become narrower,
{\em i.e.} the sensitivity increases, the more chiral structures are involved.
The pure chiral couplings prefer negative values of $\eta$.
This is a consequence of the trend of the data 
$({\rm d}\sigma / {\rm d} Q^2) \, / \, ({\rm d}\sigma^{SM} / {\rm d} Q^2)$
to be slightly low around $Q^2 \simeq 4,000 - 12,000~\GeV^2$ and being
followed by an upward fluctuation at higher $Q^2$ (see figure~\ref{cismxsec}),
which favour a negative interference term.
Note that the $LL$ and $RR$ models and the $LR$ and $RL$ models are almost
indistinguishable in deep inelastic unpolarised $e^+p$ scattering.
Within each couple the exchanged quantum numbers are the same and therefore 
the combinations $LL+RR$ and $LR+RL$ are investigated as well.
The data are more sensitive to the $VV$, $AA$ and $VA$ models,
where all chiral structures contribute. 
The most restrictive range of $\epsilon/\Lambda^2$ is obtained for the
$VV$ model, where all contact terms enter with the same sign.
Figure~\ref{cichi2eta} also shows that different parton 
distributions have little influence on the results.

\begin{table}[htb]
\begin{center}
\begin{tabular}{l c c c c  c}
   \hdick \\[-1.5ex]
   coupling & \ $\epsilon/\Lambda^2 \ [\TeV^{-2}]$ \ &
   \ $\Lambda^+~[\TeV]$ \ & \ $\Lambda^-~[\TeV]$ \ \\[1ex]
   \hdick \\[-1.5ex]
 $LL$ &$-0.249~^{+0.235+0.069}_{-0.207 -0.000}$ & 2.4 & 1.3  \\[.4em]
 $LR$ &$-0.247~^{+0.121+0.000}_{-0.064 -0.012}$ & 3.4 & 1.6  \\[.4em]
 $RL$ &$-0.226~^{+0.130+0.000}_{-0.067 -0.006}$ & 3.4 & 1.6  \\[.4em]
 $RR$ &$-0.251~^{+0.228+0.068}_{-0.209 -0.000}$ & 2.5 & 1.3  \\[.4em]
 $VV$ &$-0.028~^{+0.027+0.008}_{-0.033 -0.004}$ & 5.5 & 2.8  \\[.4em]
 $AA$ &$\phantom{-}0.131~^{+0.044+0.013}_{-0.109 -0.000}$ & 2.1 & 3.9  \\[.4em]
 $VA$ &$-0.007~^{+0.084+0.004}_{-0.071 -0.005}$ & 2.8 & 2.8  \\[.4em]
 $LL+RR$ &$-0.119~^{+0.106+0.041}_{-0.236 -0.013}$ & 3.3 & 1.4  \\[.4em]
 $LR+RL$ &$-0.046~^{+0.046+0.017}_{-0.154 -0.010}$ & 4.6 & 1.8  \\[1ex]
 \hline
\end{tabular}
\end{center}
\caption{Results of $\chi^2$ fits of the parameter $\epsilon/\Lambda^2$  
  for various chiral structures
  using CTEQ5D parton distributions with 68\%~CL intervals (first error) and
  variations due to MRST~99 and GRV~94 parton parametrisations (second error). 
  The compositeness scale parameters $\Lambda^\pm$ (95\%~CL lower limits)
  are derived as weakest bounds, {\em i.e.} smallest values
  from the analyses applying different parton densities.}
\label{etafits}
\end{table}

\begin{figure}[htb]
  \begin{center}
    \epsfig{file=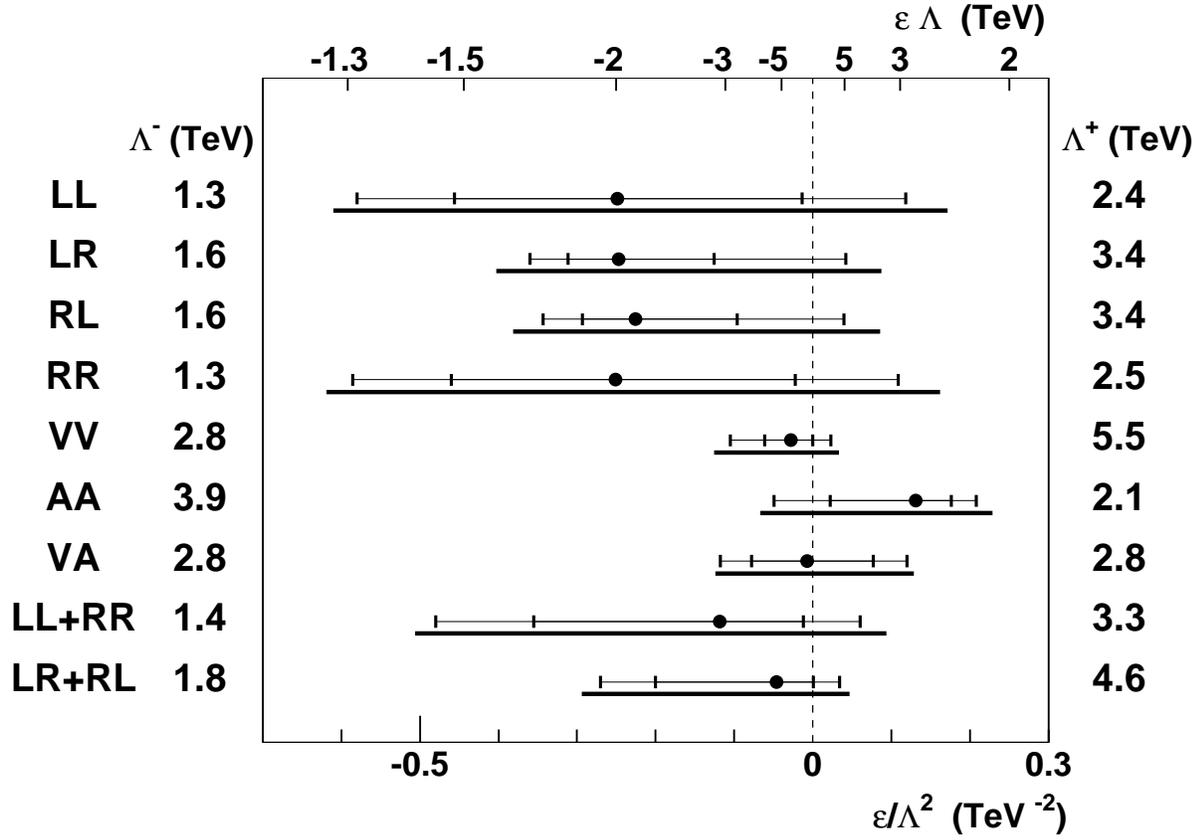,width=16cm}
  \end{center}
  \caption{Analysis results of the parameter $\epsilon/\Lambda^2$ 
    for various compositeness models.
    The thick hori\-zontal bars indicate the limits on $\Lambda^+$
    and $\Lambda^-$ including parton distribution uncertainties;
    values outside these regions are excluded at 95\% confidence level.
    The corresponding thin horizontal bars show the fit results for 
    $\epsilon/\Lambda^2$ using CTEQ5D parton distributions;
    inner and outer error bars represent one and two standard deviations
    respectively.
    The scale for $\epsilon\,\Lambda$ is shown for convenience.}
  \label{cieta}
\end{figure}

The results of the $\chi^2$ fits are shown in figure~\ref{cieta}
and compiled in table~\ref{etafits}.
Within two standard deviations the couplings $\epsilon/\Lambda^2$ are 
compatible with the Standard Model for all parton density functions used.

\begin{figure}[htb]
  \begin{center}
    \epsfig{file=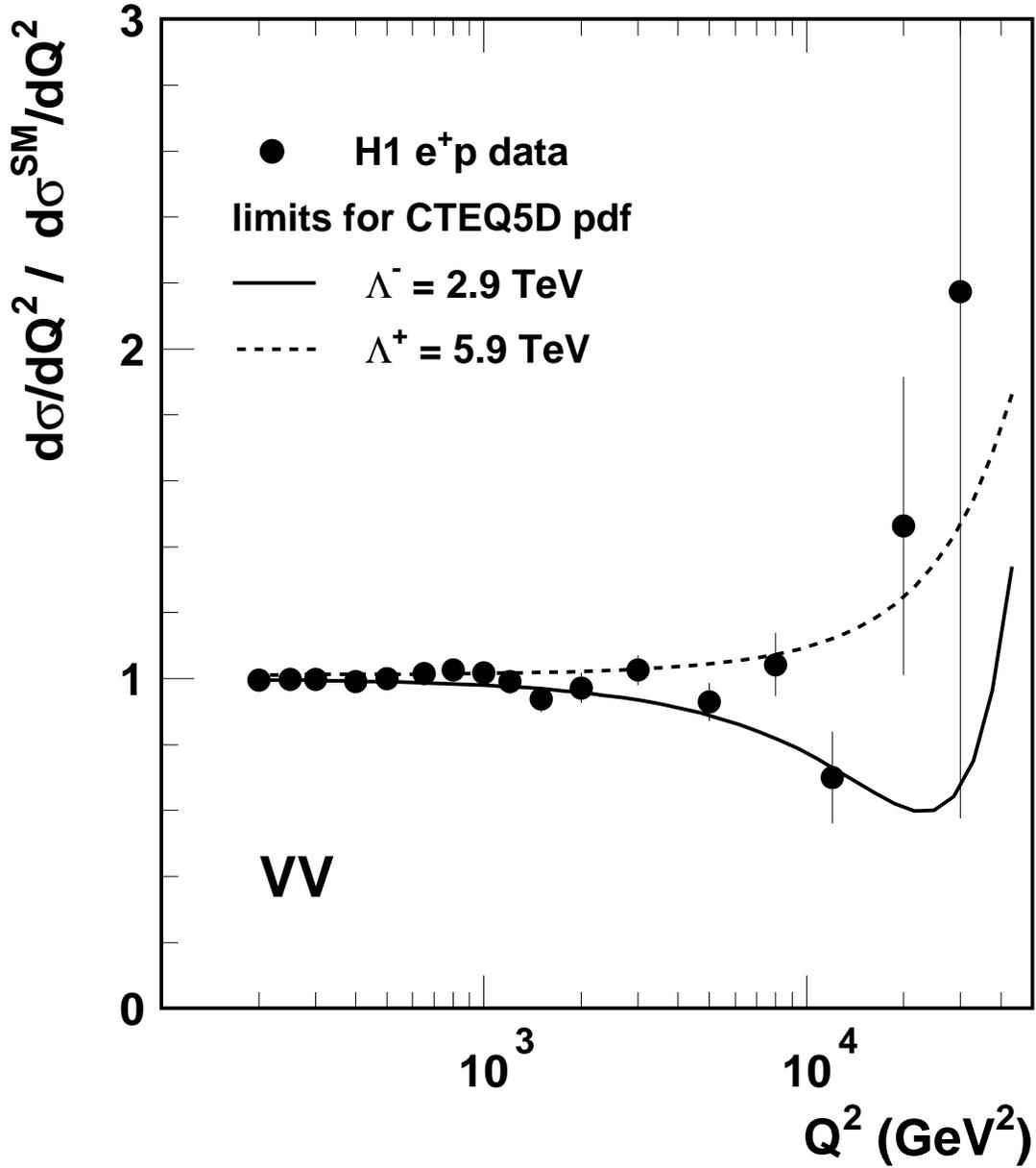,width=16cm}
  \end{center}
  \caption{NC cross section ${\rm d}\sigma / {\rm d}Q^2$ normalised to the 
    Standard Model expectation using CTEQ5D parton distributions.
    H1 data ($\bullet$) are compared with fits to the VV model corresponding
    to 95\% CL exclusion limits of $\Lambda^+$  $(- -)$ and $\Lambda^-$ (---).
    The errors represent statistics and uncorrelated experimental systematics. 
    The overall normalisation uncertainty is 1.5\%.} 
  \label{cixsec}
\end{figure}

Limits on the compositeness scale parameters $\Lambda^+$ and $\Lambda^-$, 
corresponding to positive and negative interference,
are quoted in table~\ref{etafits} and also presented in figure~\ref{cieta}.
They vary between $1.3~\TeV$ and $5.5~\TeV$ depending on the chiral structure of 
the model and are in most cases asymmetric with respect to the interference sign.
In general the values of $\Lambda^+$ are more restrictive due to the 
downward trend of the data at intermediate $Q^2$ prefering 
models with negative interference.
As an illustration of the sensitivity of the data to compositeness scales
figure~\ref{cixsec} shows the 95\%~CL contributions of $\Lambda^\pm$
for the $VV$ model using CTEQ5D parton densities.

The results of direct searches for $(e q)$ compositeness are compatible 
with those of other experiments at {\sc Hera}, {\sc Lep} and {\sc Tevatron}.
To date the most stringent limits come from $e^+e^-$ experiments~\cite{lepci} 
with typical cut-off values of $3 - 10~\TeV$ assuming, however,
in general the same scale parameter $\Lambda$ for all five active quarks.
The {\sc Zeus} collaboration~\cite{zeusci} investigates only models in which
at least 2 couplings contribute and derives limits between $1.7~\TeV$ and 
$5~\TeV$ relying solely on the shape of measured distributions.
The $p\bar{p}$ experiments~\cite{tevatronci} measure Drell-Yan production
and quote limits on $\Lambda$ between $2.5~\TeV$ and $6~\TeV$, 
where the normalisation is based on the number of observed $Z$ bosons. 
Model dependent indirect limits of order $10~\TeV$ for the pure chiral 
couplings involving $u$ and $d$ quarks can be set by atomic parity violation 
experiments~\cite{apv}.


\section{Form Factors}

An alternative method to study possible fermion substructures is to assign a
finite size of radius $R$ to the electroweak charges of leptons and/or quarks 
while treating the gauge bosons $\gamma$ and $Z$ still as pointlike 
particles~\cite{koepp}.
A convenient parametrisation is to introduce `classical' 
form factors $f(Q^2)$ at the gauge boson--fermion vertices, 
which are expected to diminish the Standard Model cross section at high 
momentum transfer
\begin{eqnarray}
  f (Q^2) & = & 1 - \frac{1}{6}\, \langle r^2 \rangle \,Q^2 \ , \\[.4em]
  \frac{d\sigma}{dQ^2} & = &
  \frac{d\sigma^{SM}}{dQ^2} \, f^2_e(Q^2)\,f^2_q(Q^2) \ .
\end{eqnarray}

The root of the mean-square radius of the electroweak charge distribution,
$R = \sqrt{\langle r^2 \rangle}$, is taken as a measure of the particle size.
The data are analysed in terms of a single form factor $f_q$,
{\em i.e.} only the quarks are allowed to be extended objects
while the positron has no structure by setting $f_e \equiv 1$.
This assumption is justified, since the pointlike nature of the 
electron/positron is already established down to extremely low distances in 
$e^+\,e^-$ and $(g - 2)_e$ experiments~\cite{kinoshita}.
The analysis yields an upper limit at 95\%~CL of the light quark radius of 
\begin{eqnarray*}
  R_q & < & 1.7 \cdot 10^{-16}~\cm \ .
\end{eqnarray*}

The result is compatible with those from other measurements.
In an analysis of Drell-Yan production of $e^+e^-$ and $\mu^+\mu^-$ pairs
in $p \bar{p}$ scattering the CDF collaboration~\cite{tevatronci} finds a 
quark size of $R_q < 1 \cdot 10^{-16}~\cm$ assuming pointlike leptons.
A complementary analysis of the contributions of anomalous magnetic 
dipole moments to the $Z q \bar{q}$ vertex using hadronic $Z$ decays gives 
$R_q < 1.2\cdot 10^{-16}~\cm$ for the light $u$ and $d$ quarks~\cite{koepp}.


\section{Leptoquarks \label{leptoquarks}}

Leptoquarks are colour triplet bosons of spin 0 or 1, carrying 
lepton ($L$) and baryon ($B$) number and fractional electric charge.
They couple to lepton--quark pairs and appear in extensions of the
Standard Model which try to establish a connection between leptons 
and quarks. 
Leptons and quarks may be arranged in common multiplets, like in Grand
Unified Theories or superstring motivated $E_6$ models, or they may have
a common substructure as in composite models.
A fermion number $F = L + 3\,B$ is defined, which takes the values
$F = 2$ for leptoquarks coupling to $e^- q$ and 
$F = 0$ for leptoquarks coupling to $e^- \bar{q}$.
For positrons the fermion number $F$ changes by two.
The leptoquark mass $M_{LQ}$ and its coupling $\lambda$ are related to the
contact interaction coefficients of eq.~(\ref{etacoeff}) via
$g/\Lambda = \lambda/M_{LQ}$. 
The notation and the coupling coefficients $\eta_{ab}$ 
for leptoquarks \footnote{The coupling coefficients are taken from 
  ref.~\cite{haberl} with the signs corrected ({\em i.e.} reversed)
  for $F=2$ scalar and $F=0$ 
  vector leptoquarks according to ref.~\cite{kalinowski}.} 
are given in table~\ref{lqfits}.
The only unknown parameter is the ratio $M_{LQ}/\lambda$.
Note that the vector leptoquarks have coupling coefficients twice as
large in magnitude compared to scalar leptoquarks. 

\begin{table}[htb]
\begin{center}
\begin{tabular}{c c c c c}
  \hdick \\[-1.5ex]
  leptoquark  & \ coupling to $u$ quark \ & \ coupling to $d$ quark \
    & \ $F$ \ & $M_{LQ}/\lambda$ \\[.5ex]
    & $[ \GeV^{-2} ]$ & $[ \GeV^{-2} ]$ & & $[ \GeV ]$ \\[1ex]
  \hdick \\[-1.5ex]
   $S_0^L$ & 
    \ $\eta^u_{LL} = +\frac{1}{2}\ (\lambda/M_{LQ})^2$ \ & & 2 & 620 \\[.2em]
   $S_0^R$ & 
    \ $\eta^u_{RR} = +\frac{1}{2}\ (\lambda/M_{LQ})^2$ \ & & 2 & 570 \\[.2em]
   $\tilde{S}_0^R$ & &
    \ $\eta^d_{RR} = +\frac{1}{2}\ (\lambda/M_{LQ})^2$ \   & 2 & 220 \\[.2em]
   $S_{1/2}^L$ &
    \ $\eta^u_{LR} = -\frac{1}{2}\ (\lambda/M_{LQ})^2$ \ & & 0 & 340 \\[.2em]
   $S_{1/2}^R$ &
    \ $\eta^u_{RL} = -\frac{1}{2}\ (\lambda/M_{LQ})^2$ \   &
    \ $\eta^d_{RL} = -\frac{1}{2}\ (\lambda/M_{LQ})^2$ \   & 0 & 320 \\[.2em]
   $\tilde{S}_{1/2}^L$ & &
    \ $\eta^d_{LR} = -\frac{1}{2}\ (\lambda/M_{LQ})^2$ \   & 0 & 450 \\[.2em]
   $S_1^L$ &
    \ $\eta^u_{LL} = +\frac{1}{2}\ (\lambda/M_{LQ})^2$ \   &
    \ $\eta^d_{LL} = +1\ (\lambda/M_{LQ})^2$               & 2 & 420 \\[1ex]
  \hline \\[-1.5ex]
   $V_0^L$ & &
    \ $\eta^d_{LL} = -1\ (\lambda/M_{LQ})^2$ \   & 0 & 670 \\[.2em]
   $V_0^R$ & & 
    \ $\eta^d_{RR} = -1\ (\lambda/M_{LQ})^2$ \   & 0 & 550 \\[.2em]
   $\tilde{V}_0^R$ & 
    \ $\eta^u_{RR} = -1\ (\lambda/M_{LQ})^2$ \ & & 0 & 410 \\[.2em]
   $V_{1/2}^L$ & &
    \ $\eta^d_{LR} = +1\ (\lambda/M_{LQ})^2$ \   & 2 & 380 \\[.2em]
   $V_{1/2}^R$ & 
    \ $\eta^u_{RL} = +1\ (\lambda/M_{LQ})^2$ \ &
    \ $\eta^d_{RL} = +1\ (\lambda/M_{LQ})^2$ \   & 2 & 960 \\[.2em]
   $\tilde{V}_{1/2}^L$ &
    \ $\eta^u_{LR} = +1\ (\lambda/M_{LQ})^2$ \ & & 2 & 1060\\[.2em]
   $V_1^L$ &
    \ $\eta^u_{LL} = -2\ (\lambda/M_{LQ})^2$ \ &
    \ $\eta^d_{LL} = -1\ (\lambda/M_{LQ})^2$ \   & 0 & 450 \\[1ex]
  \hline
\end{tabular}
\end{center}
\caption{Coupling coefficients $\eta^q_{ab}$, fermion number $F$ 
  and 95\%~CL lower limits on $M_{LQ}/\lambda$ for scalar (S) and vector (V) 
  leptoquarks, taking the most conservative values from a variation of different 
  parton distributions.
  The notation indicates the lepton chirality {\em L, R} 
  and weak isospin $I = 0,\ 1/2,\ 1$.
  $\tilde{S}$ and $\tilde{V}$ differ by two units of hypercharge from $S$ and $V$.
  By convention the quantum numbers and helicities are given for
  $e^-q$ and $e^-\bar{q}$ states.
  Limits on the coupling $\lambda$ are only meaningful for 
  leptoquark masses $M_{LQ}>\sqrt{s}$.}
\label{lqfits}
\end{table}

The differential cross section analysis gives no evidence for a 
virtual leptoquark signal.
The resulting lower limits on $M_{LQ}/\lambda$ are summarised in
table~\ref{lqfits} including the full error propagation 
and a variation of parton densities.
In general leptoquarks with positive interference provide stronger limits
compared to those with negative interference.
This observation is consistent with the results found for compositeness models. 
The vector leptoquarks which couple to $u$ quarks provide the most 
restrictive limits of $M_{LQ}/\lambda \sim {\cal O}(1~\TeV)$.
It should be emphasised that upper bounds on the coupling strength $\lambda$
can only be set for leptoquark masses exceeding the accessible center of mass
energy of {\sc Hera}.
Masses far above 300~GeV are excluded for almost all types of 
leptoquarks with a coupling of $\lambda \gtrsim 1$. 

\begin{figure}[htb]
  \begin{center}
    \epsfig{file=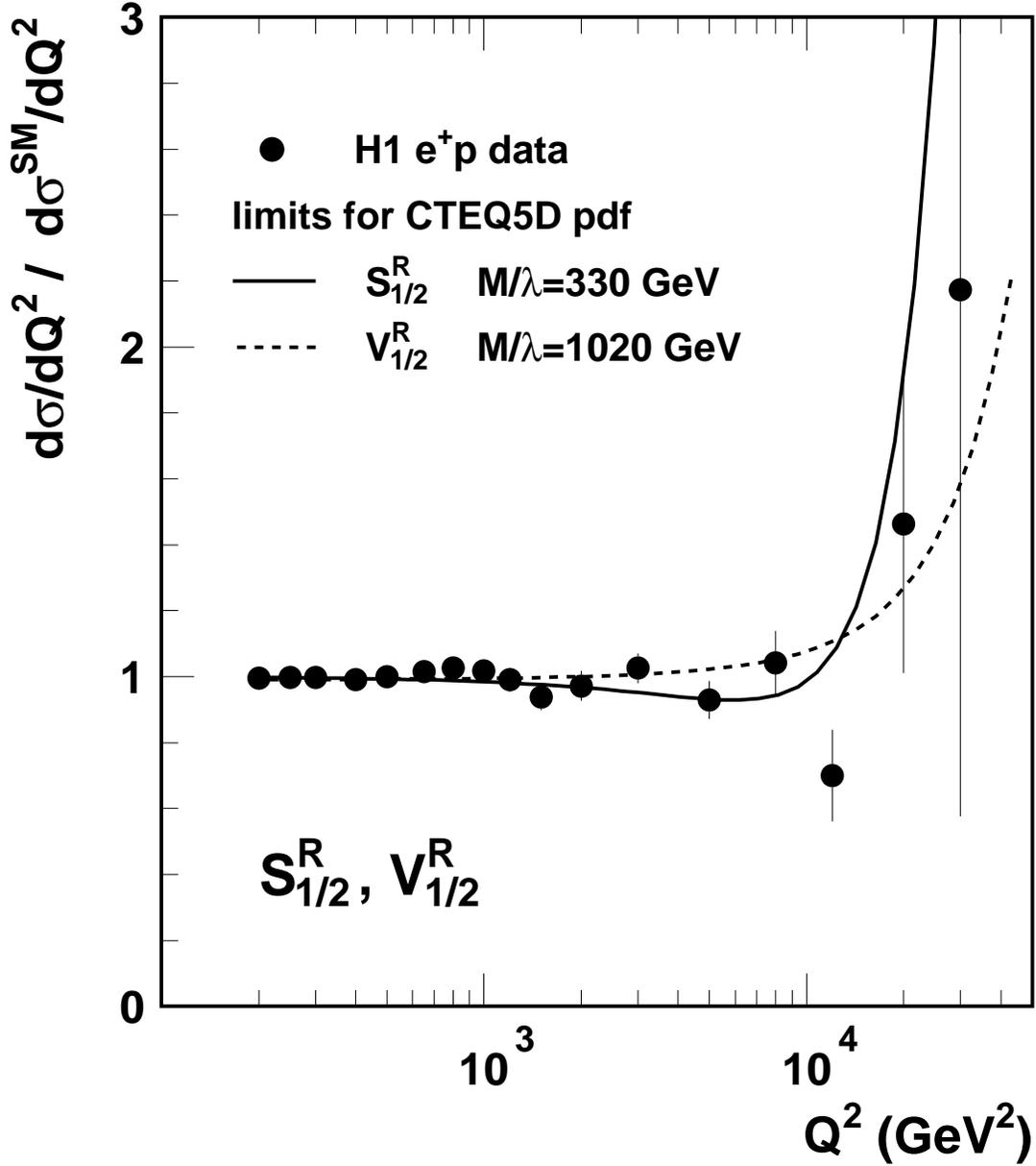,width=16cm}
  \end{center}
  \caption{NC cross section 
    ${\rm d}\sigma / {\rm d}Q^2$ normalised to the Standard Model expectation
    using CTEQ5D parton distributions.
    H1 data ($\bullet$) are compared with 95\% CL exclusion limits
    of the leptoquarks $S^R_{1/2}$  (---) and $V^R_{1/2}$  $(- -)$.
    The errors represent statistics and uncorrelated experimental systematics. 
    The overall normalisation uncertainty is 1.5\%.} 
  \label{cilqxsec}
\end{figure}

These measurements are a considerable improvement over the previous
analysis~\cite{h1ci}. 
But it should be noted that changes~\cite{kalinowski} in the signs of 
couplings $\eta_{ab}^q$ reduce the sensitivity for $F=2$ vector leptoquarks 
and lead {\em e.g.} to weaker limits for $\tilde{V}^R_0$ and $V^L_1$
despite the increased luminosity.
As an example of the sensitivity of the data to virtual leptoquark
exchange figure~\ref{cilqxsec} shows the contributions given by the lower
limits on  $M_{LQ}/\lambda$ for the scalar leptoquark $S^R_{1/2}$ 
and the vector leptoquark $V^R_{1/2}$.
Both leptoquarks have $RL$ couplings to {\em up} and {\em down} quarks,
which differ in magnitude and interference sign.

The present contact interaction results complement the direct leptoquark
searches of the H1 collaboration~\cite{h1lq},
which have recently been extended beyond the kinematic reach of {\sc Hera}
up to masses of $M_{LQ} \simeq 400~\GeV$. 
The coupling limits derived in both analyses are compatible with each other
in the mass region where they overlap.
Virtual leptoquark exchange has also been studied in $e^+e^-$ 
annihilation experiments at {\sc Lep}~\cite{lepci}. 
Typical limits on $M_{LQ}/\lambda$ are in the range $0.3 - 1.8~\TeV$, 
but the sensitivity to particular leptoquark types is different from
deep inelastic $e^+p$ scattering. 
In most cases the {\sc Lep} results provide more stringent bounds;
the limits for $S^R_0$, $\tilde{S}^L_{1/2}$ and $V^R_0$ are comparable
and those of $V^R_{1/2}$ and $\tilde{V}^L_{1/2}$ are superior at {\sc Hera}.


\section{Large Extra Dimensions}

The contributions of virtual graviton exchange to deep inelastic scattering
have been derived from the cross sections given in ref.~\cite{giudice}
for $e^+e^-$ collisions by applying crossing relations.
The basic processes of elastic $e^+q \to e^+q$ and $e^+g \to e^+g$ scattering 
can be written as \footnote{
  The following formulae of ref.~\cite{giudice} are used: 
  eq.~(79) for the $e q$ contributions,
  eq.~(77) for the $e g$ contribution replacing the photons by gluons,
  and eqs.~(A.5), (A.7) -- (A.9) to expand the functions $G_i(s/t)$.
  The present $e q$ results are in agreement with \cite{cheung}, 
  but the $e g$ part differs by a factor of $1/4$.
  The cross section formulae of 
  refs.~\cite{mrs} and \cite{rizzo} cannot be confirmed and the results of
  both publications are inconsistent with each other.}
\begin{eqnarray}
  \frac{\mathrm d \sigma (e^+q \rightarrow e^+q)}{\mathrm d t} & = &
  \frac{\mathrm d \sigma^{SM}}{\mathrm d t} +
  \frac{\mathrm d \sigma^{G}} {\mathrm d t} +
  \frac{\mathrm d \sigma^{\gamma G}}{\mathrm d t} +
  \frac{\mathrm d \sigma^{Z G}}{\mathrm d t} \ , \\[.4em]
  \frac{\mathrm d \sigma^{G}}{\mathrm d t} & = &
  \frac {\pi\, \lambda^2}{32\, M_S^8} \, \frac{1}{s^2} 
  \left\{32\,u^4 + 64\,u^3 t + 42\,u^2 t^2 + 10\,u\,t^3 + t^4\right\} \ , \\[.4em]
  \frac{\mathrm d \sigma^{\gamma G}}{\mathrm d t} & = &
  -\frac{\pi\, \lambda}{2\, M_S^4} \,
   \frac{\alpha\, e_q}{s^2} \, \frac{(2\, u + t)^3}{t} \ , \\[.4em]
  \frac{\mathrm d \sigma^{Z G}}{\mathrm d t} & = &
  \frac{\pi\, \lambda}{2\, M_S^4} \,\frac{\alpha}{s^2\sin^2 2\,\theta_W} 
  \left \{ v_e v_q\, \frac{(2\,u + t)^3}{t-m_Z^2} 
         - a_e a_q\, \frac{t\,(6\,u^2 + 6\,u\,t + t^2)}{t-m_Z^2} \right \} 
      \, , \ \ \ \\[.4em]
  \frac{\mathrm d \sigma(e^+g \rightarrow e^+ g)}{\mathrm d t} & = &
  \frac{\pi\, \lambda^2}{8\, M_S^8} \, \frac{u}{s^2} 
  \left\{2\, u^3 + 4\, u^2 t + 3\, u\, t^2 + t^3 \right\} 
\end{eqnarray}
in an obvious notation of Standard Model (SM),
pure graviton (G), $\gamma G$ and $Z G$ interference contributions.
Here $s$, $t = - Q^2$ and $u$ are the Mandelstam variables,
$e_q$ is the quark charge and $v_f$ and $a_f$ are the vector and axial vector 
couplings of the fermions to the $Z$.
The corresponding cross sections for $e^+ \bar{q}$ scattering are obtained
by replacing $e_q \to -e_q$ and
$v_q \to -v_q$ in the expressions above.
In order to get the inclusive $e^+p$ cross section the subprocesses have to
be integrated over the  $x$ dependent parton distributions,
$q(x)$, $\bar{q}(x)$ and $g(x)$, and to be summed up
\begin{eqnarray}
\frac{\mathrm d\sigma (e^+ p \rightarrow e^+ X)}{\mathrm d Q^2} & = &
   \int \!\mathrm d x \, \left \{
   q(x)\, \frac{\mathrm d \sigma(e^+q)}{\mathrm d t} 
   + \bar{q}(x)\, \frac{\mathrm d \sigma(e^+\bar{q})}{\mathrm d t} 
   + g(x)\, \frac{\mathrm d \sigma(e^+g)}{\mathrm d t}  \right \} \ .
\end{eqnarray}
Note, that expected gravitational effects arising from the gluon contribution are 
for the highest $Q^2$ values of the order percent
compared to those coming from the quarks and antiquarks.

\begin{figure}[hbt]
  \begin{center}
    \epsfig{file=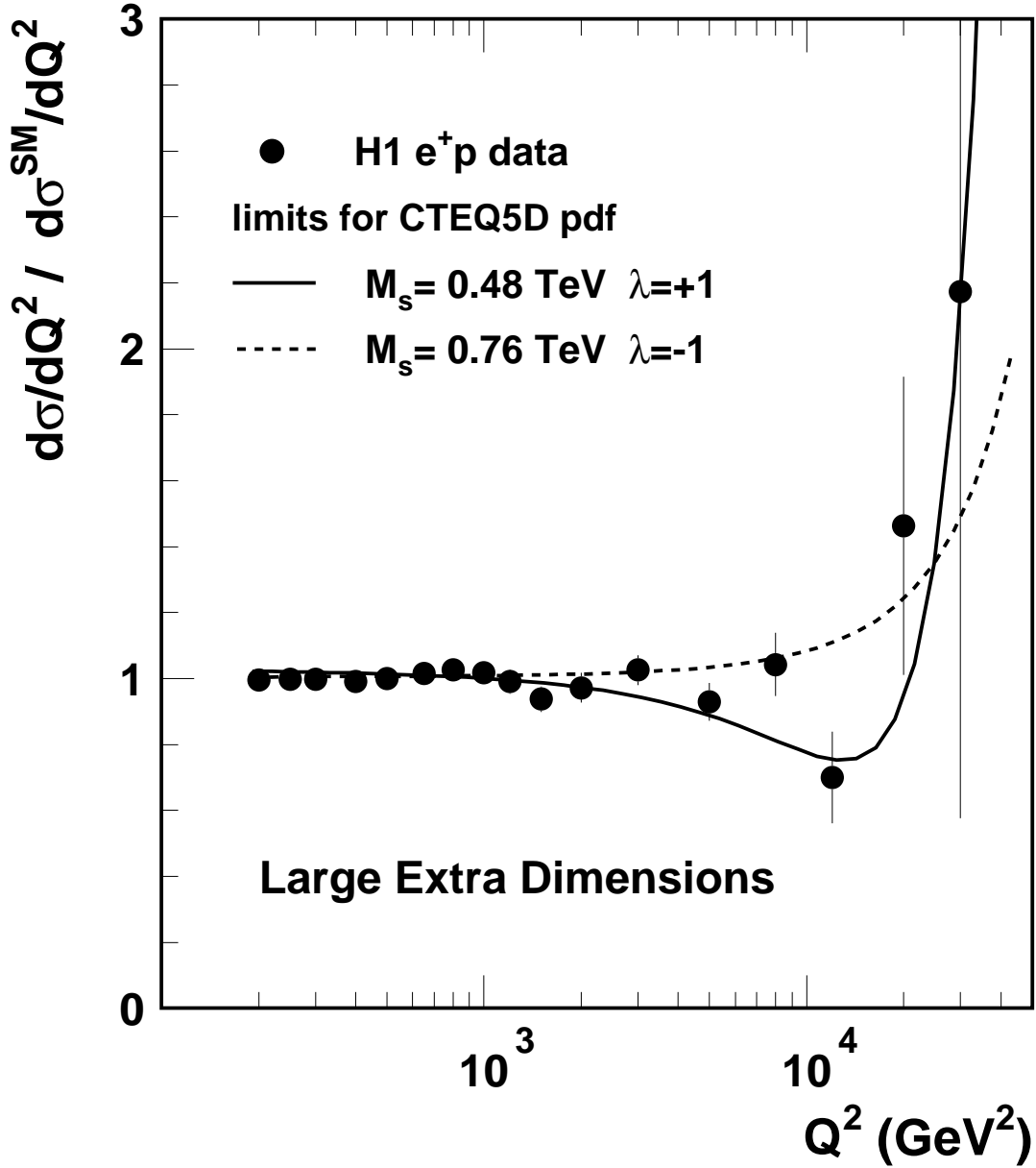,width=16cm}
  \end{center}
  \caption{NC cross section 
    ${\rm d}\sigma / {\rm d}Q^2$ normalised to the Standard Model expectation
    using CTEQ5D parton distributions.
    H1 data ($\bullet$) are compared to the effect of graviton exchange given
    by the lower limits (95\% confidence level) on the scale $M_S$ for positive 
    ($\lambda = +1$, full curve) and negative ($\lambda = -1$, dashed curve) 
    coupling.
    The errors represent statistics and uncorrelated experimental systematics. 
    The overall normalisation uncertainty is 1.5\%.} 
\label{graveffect}
\end{figure}

The strength of virtual graviton exchange to the cross section contributions
is characterised by the ratio $\lambda/M^4_S$.
The coupling $\lambda$ depends on the full theory and is expected to be of 
order unity.
Also the sign of interference with the Standard Model particles is 
{\em a priori} not known.
Therefore the coupling is set to $\lambda = \pm 1$, following the convention 
of \cite{giudice}.
The data analysis is similar to the procedures described above.
Gravitational effects are searched for by fitting the differential cross 
section to the above formulae with $\lambda/M^4_S$ treated as free parameter.
The result of $\lambda/M^4_S = 3.3~^{+4.2}_{-3.3}~^{+0.4}_{-1.3}~\TeV^{-4}$, 
where the second error reflects the parton density variation,
is compatible with the Standard Model expectation.
Lower limits at 95\%~CL on $M_S$ for positive and negative 
coupling are then derived from the change in $\chi^2$ with respect to the 
Standard Model fit, yielding 
\begin{eqnarray*}
  M_S > 0.48~\TeV & & {\rm for} \ \lambda = +1 \ , \\
  M_S > 0.72~\TeV & & {\rm for} \ \lambda = -1 \ .
\end{eqnarray*}
Possible effects of graviton exchange to the data, 
as given by the exclusion limits, are illustrated in figure~\ref{graveffect}.

Similar investigations of virtual graviton effects in $e^+e^-$ annihilation
into fermion and boson pairs provide comparable limits~\cite{leped}.
From the corresponding reaction of quark pair production scales of $M_S$ 
lower than $0.5 - 0.65~\TeV$ can be excluded. 
Combining all reactions that lead to two-fermion final states limits approaching
$1~\TeV$ can be set.


\section{Conclusions}

Neutral current deep inelastic cross section measurements
are analysed to search for new phenomena mediated through 
$(\bar{e} e)(\bar{q} q)$ contact interactions.
No significant signal for compositeness, a quark form factor and virtual 
leptoquark or graviton exchange is found and the data are used to set limits
which supersede and substantially improve former H1 results~\cite{h1ci}.

Limits on $(e q)$ compositeness 
are derived within a model independent analysis for scenarios involving 
one or more chiral couplings.
The lower bounds on the scale parameters $\Lambda^\pm$ range between 
$1.3~\TeV$ and $5.5~\TeV$ for a coupling strength $g = \sqrt{4\,\pi}$, 
depending on the chiral structure of the model.

A different approach to substructures is the concept of form factors.
Such an analysis yields an upper limit of the size of the light
$up$ and $down$ quarks of $R_q < 1.7\cdot 10^{-16}~\cm$ assuming
a pointlike lepton.

A study of virtual leptoquark exchange yields lower limits on 
the ratio $M_{LQ}/\lambda$ which for all types (except one) exceed the collider
center of mass energy and approach $1~\TeV$ for vector leptoquarks with 
couplings to {\em up} quarks.
These measurements complement and extend the direct leptoquark searches at 
{\sc Hera} to high masses $M_{LQ} > \sqrt{s}$.

In a search for
possible effects of low scale quantum gravity with gravitons coupling to
Standard Model particles and propagating into extra spatial dimensions,
lower limits on the effective Planck scale $M_S$ of $0.48~\TeV$ and
$0.72~\TeV$ for positive and negative coupling, respectively, are found.

\bigskip\bigskip\noindent
{\bf Acknowledgements}. 
We are grateful to the {\sc Hera} machine group whose outstanding 
efforts have made and continue to make this experiment possible. We thank 
the engineers and technicians for their work in constructing and now 
maintaining the H1 detector, our funding agencies for financial support, the 
{\sc Desy} technical staff for continual assistance, and the 
{\sc Desy} directorate for 
the hospitality extended to the non--{\sc Desy} members of the collaboration. 
We gratefully \mbox{acknowledge} valuable discussions with G.F.~Giudice.

%
%

\clearpage
%
%

%

%

%

%

\end{document}